\documentclass[runningheads]{llncs}
\usepackage{graphicx}
\usepackage{psfrag}
\usepackage{url}

\newcommand{\journal}[1]{\relax}
\newcommand{\conferenceonly}[1]{#1}

\newtheorem{thm}{Theorem}

\def\figname{Fig.~}
\def\bez{B\'ezier{} }

\begin{document}

\title{Confluent Layered Drawings
\thanks{%
Work by the first author is supported by NSF grant CCR-9912338. 
Work by the second and the third author is supported by 
NSF grants CCR-0098068, CCR-0225642, and DUE-0231467.}%
}

\titlerunning{Confluent Layered Drawings}

\author{David~Eppstein%
\and Michael~T.~Goodrich%
\and Jeremy~Yu~Meng%
}

\authorrunning{Eppstein, Goodrich, and Meng}


\institute{
School of Information and Computer Science,\\
University of California, Irvine,\\
Irvine, CA 92697, USA\\
\email{\{eppstein, goodrich, ymeng\}@ics.uci.edu}
}

\maketitle

\begin{abstract}
We combine the idea of confluent drawings with Sugiyama style
drawings, in order to reduce the edge crossings in the resultant
drawings.  Furthermore, 
it is easier to understand the structures of graphs
from the mixed style drawings.  
The basic idea is 
to cover a layered graph by complete
bipartite subgraphs (bicliques),
then replace bicliques with tree-like structures.  The
biclique cover problem is reduced to a special edge
coloring problem and solved by heuristic coloring algorithms.
Our method can be extended to obtain multi-depth confluent layered 
drawings.
\end{abstract}

\section{Introduction}
\label{sec:intro}

Layered drawings visualize hierarchical graphs in a way such that
vertices are arranged in layers and edges are drawn 
as straight lines or curves connecting these layers.
A common method was introduced by Sugiyama, Tagawa and 
Toda~\cite{stt-mvuhs-81} and by Carpano~\cite{c-adhgc-80}.  
Several closely related methods were proposed later 
(see e.g.~\cite{gnv-dptdd-88,m-alldg-88,gm-radhd-89,%
nt-eege-90,es-hddg-91,mrh-dcaal-91,gknv-tddg-93}.)  

Crossing reduction is one of the most important objectives in 
layered drawings.  But it is well known that 
for two-layer graphs 
the straight-line crossing minimization problem 
is NP-complete~\cite{gj-cninc-83}.
The problem remains NP-complete even when one layer is fixed.
J{\"u}nger and Mutzel~\cite{jm-2lscm-97} 
present exact algorithms for this problem, 
and perform experimental comparison of their results  
with various heuristic methods. 
\journal{
They conclude that 
if the permutation of one layer is fixed, 
there is no need for heuristics, and in the case where 
two layers are not fixed, 
the iterated barycenter method is 
the best among several heuristics.
}
Recently new methods related to crossing reduction
(\cite{wm-elelc-99,bjm-sebcc-02,egdb-crwo-02,nsv-tnhts-02,f-acrsl-02})
have been proposed.

However when the given two-layer graph is dense, 
even in an optimum solution, 
there are still a large number of 
crossings.  
Then the resulting straight-line drawing will be hard to read, since 
edge-crossing minimization is one of the most 
important aesthetic criteria for visualizing graphs~\cite{p-wageh-97}.
This give us a motivation for exploring new approaches to 
reduce the crossings in a drawing other than the traditional methods.

In addition, it is sometime of interest to 
find the bicliques
between two layers.  For example in the drawing
of a call graph, 
it is interesting
to find out 
which set of modules are calling 
a common set of functions and what are those common
functions.  
Call graphs are usually visualized as layered drawings.
However it is hard to learn this information from
layered drawings by traditional Sugiyama-style approaches, 
especially when the input graphs are dense.

Our previous work~\cite{degm-cdvnd-03} introduces the concept of 
confluent drawings.  
In~\cite{degm-cdvnd-03} we talk about the confluent drawability
of several classes of graphs and 
give a heuristic for finding confluent drawings
of graphs with bounded arboricity.
In this paper we experiment with
an implementation of confluent drawings
for the layered graphs.
However we relax the constraint of planarity and 
allow crossings in the drawings,
while it is not allowed to have crossings 
in a confluent drawing in our previous definitions.

We are aware of the Edge Concentration method by
Newbery~\cite{n-ecmcd-89}.  Edge Concentration and our method 
share a same idea of covering by
bicliques.
But in Newbery's method, dummy nodes
(edge concentrators) are explicit in the drawing and treated equally
as original nodes, 
which causes the nodes' original levels to change.
In our
method dummy nodes are implicit in the curve representation of
edges and the original levels are preserved.
Furthermore, our method uses a very different algorithm to compute the
biclique covers.

\section{Definitions}
\label{sec:definitions}

\begin{figure}[htbp]
\vspace*{-7pt}
  \centering
  \includegraphics[width=.42\textwidth]{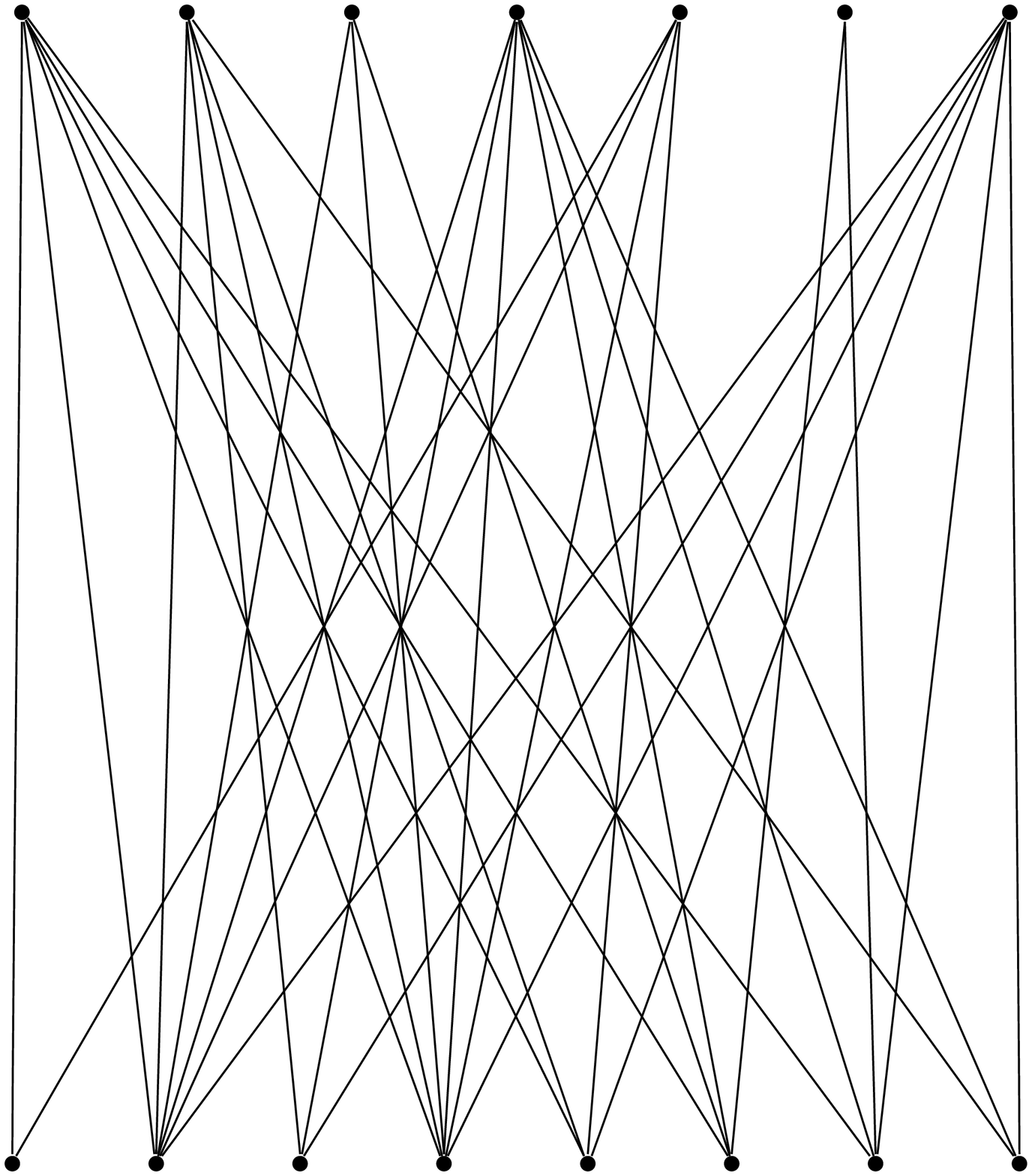}\hfil
  \includegraphics[width=.42\textwidth]{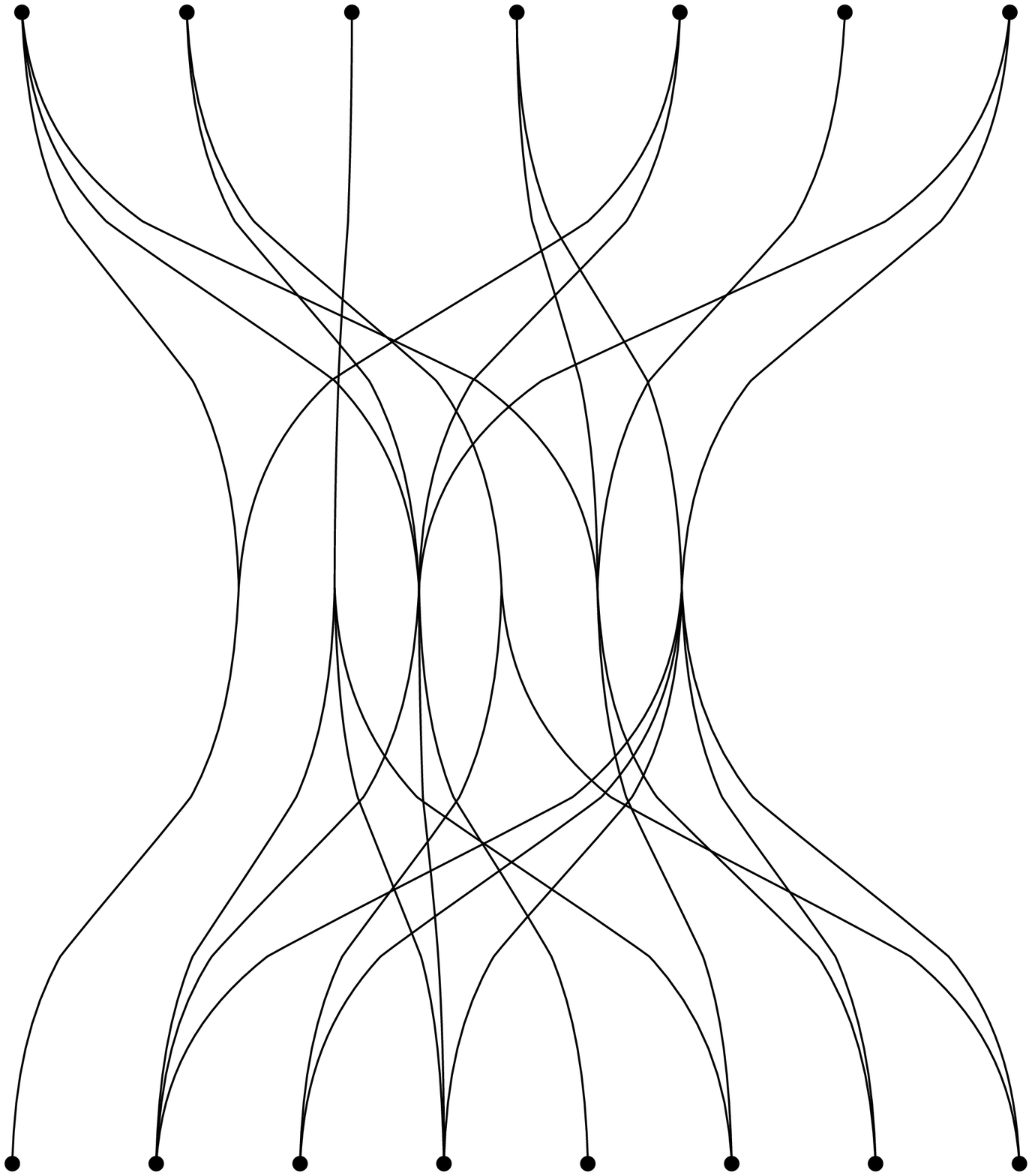}
\vspace*{-7pt}
  \caption{An example confluent layered drawing.} 
  \label{fig:first} 
\end{figure}

In this section we give definitions for confluent layered drawings.
The definitions almost remain the same as in our previous
confluent drawing paper, 
except that the planarity
constraints are dropped. 
\figname\ref{fig:first} gives an idea of confluent layered drawings.
Edges in the drawing are represented as smooth curves.

A curve is \textit{locally monotone\/} if it contains  
no sharp turns, that is, it contains no point with
left and right tangents that forms a angle less than or equal to 90
degrees.  Intuitively, a locally-monotone curve is like a single train
track, which can make no sharp turns.  
\textit{Tracks\/} are the union
of locally-monotone curves. 
They are formed by merging edges together.

A drawing $A$ formed by a collection of tracks 
on the plane is called a \textit{confluent drawing\/} 
for an undirected graph $G$ 
if and only if 

\begin{itemize}
\setlength{\itemsep}{0pt}
\setlength{\parsep}{0pt}
\item 
  There is a one-to-one mapping between the vertices in $G$ and
  $A$, so that, for each vertex $v \in V(G)$, there is a corresponding vertex 
  $v' \in A$, 
  and all vertices of $G$ are assigned to distinct points in the plane.
\item 
  There is an edge $(v_i,v_j)$ in $E(G)$
  if and only if there is a locally-monotone curve $e'$
  connecting $v_i'$ and $v_j'$ in along tracks in~$A$.  
\end{itemize}

The directed version 
of a confluent drawing is defined similarly, except
that in such a drawing the locally-monotone curves are directed and
in every track formed by the union of directed curves, the curves must be
oriented consistently. 

Self loops and parallel edges of $G$
are not allowed in our definitions,
although multiple ways of realizing 
the same edge are allowed.  
Namely, for
an edge in the original graph, 
there could be more than one 
locally monotone path
in the drawing
corresponding to this edge.

We apply 
the idea of confluent drawings on layered graphs. 
Particularly, in the resultant confluent drawing,
we replace bicliques in a biclique cover of 
a two-layer graph $G=(U,L,E)$ by tree-like structures 
and draw them with smooth curves. 
As we can see in \figname\ref{fig:first}, 
our method can greatly reduce the crossings in the drawings of 
dense bipartite graphs.
Additionally, 
nodes of a biclique can be easily 
identified by following the smooth curve paths.   

Since it
is valid to have more than one confluent 
path 
between two nodes $u$ and $l$ in the confluent drawing when
$(u,l)\in E$, as 
defined above,
it is straightforward that
a confluent layered drawing can be obtained 
 by computing a biclique cover $C$ of $G$,
then visualizing each biclique in $C$ as a tree-like structure. 
We show how to compute a biclique cover of $G$ in the next section.

\section{Computing Biclique Cover of Bipartite Graphs}
\label{sec:comp-bicl}

Fishburn and Hammer~\cite{fh-bdbdg-96} show that the biclique cover
problem is equivalent to a restricted edge coloring problem.  
This coloring is not much useful for general
graphs. However, it has a nice result for 
triangle-free graphs, and since bipartite graphs belong to 
the class of triangle-free
graphs, an immediate result is that this 
type of edge coloring can be 
used to find a 
biclique cover of a bipartite graph.  
This result is useful in layered drawing because 
the edges between any two layers in such a drawing
induce a bipartite subgraph.

\begin{figure}[htb]
  \psfrag{c1}{$c_1$} \psfrag{c2}{$c_2$} \psfrag{c3}{$c_3$}
  \psfrag{P_4}{induced $P_4$} \psfrag{C4c}{induced $C_4^c$}
  \psfrag{cond1}{$|\{c_1,c_2,c_3\}| \ge 2$}
  \psfrag{cond2}{$c_1 \neq c_2$}
\vspace*{-7pt}
  \centering
  \includegraphics[width=0.6\textwidth]{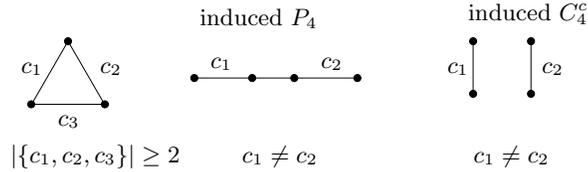}
\vspace*{-7pt}
  \caption{The required conditions of
    induced subgraphs of a simply edge coloring.}
  \label{fig:forbidden}
\end{figure}

An edge coloring 
$c\colon E \leftarrow \{1,2,\ldots,k\}$ for $G=(V,E)$ 
is \textit{simply-restricted\/} if no induced $K_3$ is monochromatic 
and the vertex-disjoint edges in an induced $P_4$ or $C_4^c$ have
different colors.  
\figname\ref{fig:forbidden} shows the conditions 
that such induced subgraphs of a 
simply-restricted edge 
coloring must satisfy.

Let $d(G)$ denote the bipartite dimension of $G$, which is the minimum
cardinality of a biclique cover of $G$.  Let $\chi_s(G)$ be the
chromatic number of a 
simply-restricted edge coloring of $G$.  
$\chi_s(G)$ is
$0$ if $E=\emptyset$; otherwise, it is 
the minimum $k$ for which $G$
has a simply-restricted coloring $c\colon E \leftarrow \{1,2,\ldots,k\}$.
The following theorem states 
the equivalence of $d(G)$ and $\chi_s(G)$
for triangle-free graphs.

\begin{thm}
\textbf{(Fishburn and Hammer~\cite{fh-bdbdg-96})}

$d(G)=\chi_s(G)$ for every triangle-free graph.
\label{thm:k3free}
\end{thm}

\journal{
For completeness, we include the proof of Theorem~\ref{thm:k3free} 
in Appendix~\ref{sec:appx-1}.  
The proof is also helpful for readers to 
understand the algorithm.
}

Let $E_j$ be the set of edges with color $j$ in 
a simply-restricted edge coloring for a
triangle-free graph $G$.  
As we can see in the second part of the proof of 
Theorem~\ref{thm:k3free} 
(omitted here, included in the full version of this paper), 
$E_j$ is included in the edge set of a
biclique subgraph of $G$.  
Therefore, 
every edge set of a 
single color  
induces a biclique subgraph of $G$.  
By computing a simply restricted edge coloring 
we can get a biclique cover of $G$.

Because it is known that the problem of 
\textsc{Covering by Complete Bipartite Subgraphs}
is NP-hard (Garey and Johnson~\cite{gj-cigtn-79} GT18), 
it is unlikely to have efficient optimization algorithms for 
finding the minimum biclique cover of a bipartite
graph. 
Thus we only focus on fast heuristics for 
computing a near-optimal biclique cover. 

\journal{
Note that 
given
$G=(V,E)$, 
we can construct a new graph $G_E=(V',E')$ such that,
for every edge $e\in E$ there is a vertex in $V'$, and there is an edge
in $E'$ between two vertices in $V'$ if their corresponding $e$ and $e'$
are vertex-disjoint and the two lie in an induced $P_4$ or $C_4^c$. 
If $G$ is triangle-free, then $d(G)$ is the vertex chromatic number of
$G_E$.
This transformation from simply-restricted edge coloring 
into vertex coloring makes 
it possible to use vertex coloring 
algorithms to compute biclique covers. 
}
\conferenceonly{
The simply-restricted edge coloring problem can be transformed
into a vertex coloring problem.  
}
So, instead of devising a special algorithm 
for the simply-restricted edge coloring, we can choose to use 
one of the existing vertex coloring algorithms.
Well known heuristic algorithms for vertex coloring include 
Recursive Largest First (RLF) algorithm of 
Leighton~\cite{l-gcals-79},
DSATUR algorithm of Br\'elaz~\cite{b-nmcvg-79}.
For more about heuristics on graph coloring, 
see Campers et al.~\cite{chl-gchss-88}. 

\conferenceonly{The above method of 
computing a biclique cover by coloring doesn't distinguish between
two kinds of bicliques: $K_{p,1}$ and $K_{1,r}$, where $p,q,r>1$. 
So if we are more interested in finding out 
the set of common callers and callees,  
we would need to give higher priority to $K_{p,q}$ than $K_{1,r}$
when covering the edges.
}
\journal{
There is one thing worth noting: this method of 
computing a biclique cover by vertex coloring 
doesn't distinguish between two kinds of bicliques:
bicliques with at least two nodes
in each partition of its vertex set,
and 
bicliques having at least two nodes 
in only one partition of its vertex set.
So if we are more interested in finding out 
the set of common callers and callees
as mentioned in the introduction, we would want 
to give higher priority to $K_{p,q}$
than $K_{1,r}$ (where $p,q,r>1$)
when covering the edges.  
This can be done in the coloring process 
by picking first the 
nodes of $G_E$ that represent vertex-disjoint edges in 
the original bipartite graph~$G$.  
But then we would need to pass extra information to the 
coloring procedure, because $G_E$ doesn't contain
information about whether two edges 
of $G$ are sharing a vertex or not. 
It has only the information of whether 
two edges could be colored using
same color.
}

After the biclique cover of 
the two-layer bipartite graph is computed, each 
biclique in the cover 
is drawn 
as a tree-like structure in the
final drawing.  
Doing this repeatedly for every two adjacent layers,
we can get the drawings for
multi-layer graphs.

The time complexity of the algorithm depends on the coloring
heuristic subroutines.  
For a graph with a set of vertices $V$,
both the RLF algorithm and the DSATUR
algorithm run in worst case $O(|V|^3)$ time.  There are some other
faster coloring heuristics with $O(|V|^2)$ time, 
but their output qualities are worse.
Suppose we have a two-layer bipartite graph
$G=(V,E)$.  
The transformation from the simply-restricted edge coloring
into vertex coloring version takes $O(\left|E\right|^2)$ time.  Using RLF
or DSATUR costs $O(|E|^3)$,  thus the total time is $O(|E|^3)$.

\section{Layout of the bicliques}
\label{sec:placement}

We described how to compute a biclique cover of 
a two-layer bipartite graph in the previous section.  
Now it is time to show how 
the bicliques are laid out.
In the confluent layered drawings,
each biclique in the biclique cover is
visualized as a tree-like structure, as 
in \figname\ref{fig:first}.  Now here are the questions.
What are the best positions to place 
the centers of the tree-like structures?
How to arrange the curves so that they 
form confluent tracks defined in Section~\ref{sec:definitions}?

\subsection{Barycenter method to place centers}

\begin{figure}[htb]
\vspace*{-7pt}
  \centering
  \includegraphics[scale=.7]{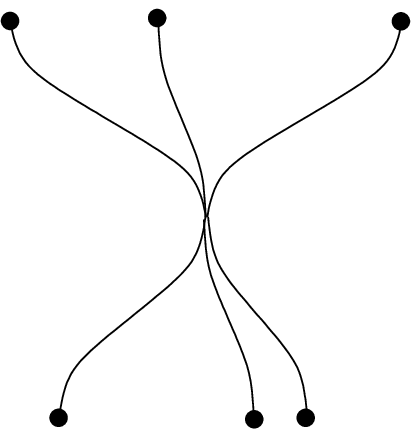}%
  \hspace{40pt}%
  \includegraphics[scale=.7]{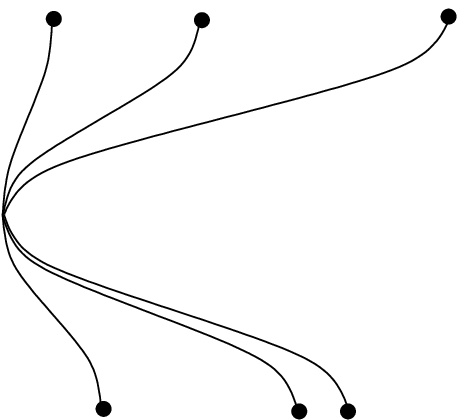}
\vspace*{-7pt}
  \caption{Good-looking tree and bad-looking tree with centers placed differently.}
  \label{fig:badcenter}
\end{figure}

In the case where the positions of nodes in the upper level 
and lower level are fixed, 
one would like to put the center of a tree to 
the center of the nodes belonging to 
the corresponding biclique.  
For example, in \figname\ref{fig:badcenter}, 
the drawing on the left is 
visually better than the drawing on the right. 
Firstly it has better angular resolution and 
better edge separation.  
Secondly it is easier for people to see the biclique as a whole. 
Then the next question is:
what does the center of those nodes mean?  
In our method, the natural candidate position for a center of 
the tree-like structure is the barycenter, 
i.e., the average position, 
of all the nodes in this biclique.

It looks bad too if these tree centers stay very 
close to each other.
So we need to specify a minimum separation between 
two centers.

The above requirements can be formulated into 
constraints:
\vspace*{-7pt}
\begin{enumerate}
\setlength{\itemsep}{0pt}
\setlength{\parsep}{0pt}
\item A tree center stays within the range of its leaves.
\[
\mathop{\min}\limits_j x_{ij}%
\leq x_{i}%
\leq \mathop{\max}\limits_j x_{ij},
\]

where $x_i$ is the $x$-coordinate of the $i^{\rm th}$ 
tree center $c_i$, 
and $x_{ij}$ is the $x$-coordinate of 
the $j^{\rm th}$ leaf of $c_i$. 

\item The distance between any two centers is greater than 
or equal to the 
minimum separation.  
\[
\forall\, i \ne j,\qquad |x_{i}-x_{j}| \geq \delta
\]

where $\delta$ is some pre-specified minimum distance.
\end{enumerate}

Under these constraints, we want a tree center to stay 
as close as possible to the barycenter of all its leaf nodes.
i.e., we want to minimize 
$\sum_{i}(x_{i}-{\rm avg}_j(x_{ij}))^2$, 
subject to the above 
constraints.  
This is a Quadratic Programming problem, and unfortunately 
it is NP-hard (Garey and~Johnson~\cite{gj-cigtn-79},~MP2). 

Since it is unlikely to 
have efficient algorithms for solving
this optimization problem, 
and a small deviation of a tree center from the 
perfect position won't cause too much displeasure,
we use instead a very simple heuristic method to 
place the tree centers.  
We first assign 
to each tree center the $x$-coordinate of the  
barycenter of its leaves.  
Then we sort tree centers by their $x$-value.  
The third step is
to try to place these tree centers at their $x$-coordinates 
one by one.
Assume we have $k$ centers to place.
Start from the $j^{\rm th}$ center,
where $j=\lfloor\frac{k}{2}\rfloor$.
Place center $j$ at its barycenter, then try to 
place centers one by one in the following order:~$j-1,j-2,\ldots,1$. 
If constraint $2$ is violated, the violating center is 
placed the minimum distance away from the previous placed center.
Tree centers to the right of center $j$ are placed similarly 
in the order of~$j+1,j+2,\ldots,k$. 
It is easy to see that 
the running time of the barycenter method is dominated 
by the sorting of the tree centers.

\subsection{Placing tree centers to reduce crossings}
\label{sec:redu-xings}

Alternatively, one might want to place these centers 
on positions such
that the total number of edge crossings is as few as possible,
especially in the case 
where nodes of upper level and lower level are not fixed.
If this is the main concern, 
we can place the tree centers in another way 
in order to reduce the edge crossings.

After the biclique cover of a two-layer graph 
$G=(U,L,E)$ is computed, we construct a new three-layer graph $G'$. 
We treat these tree centers as nodes of a middle layer.
The set of vertices includes three levels: 
an upper layer $U'=U$,
a middle layer~$M$ consisting of tree centers, 
and a lower layer~$L'=L$.
The edges of $G'$ are added as follows:
for each biclique $B_i$ in the biclique cover, 
add one edge between the tree center node $m_i$ 
and each node $u\in U$ that belongs to $B_i$.
Similarly add one edge between $m_i$ and each node $l\in L$
that belongs to $B_i$.

Now a two-layer graph of the original problem is 
transformed into a three-layer graph $G'$.  
Straight-line crossing reduction algorithms can be
applied on $G'$.  
After the crossing reduction, we obtain the ordering of nodes
in each of the three layers.  
The orderings will be used to compute the positions of nodes 
and tree centers in the final drawings.  
Note that when crossing reduction method is used to place tree centers,
it is not always true that 
a tree center always stays within the $x$-range of its leaves,
i.e., bad centers like the one in \figname\ref{fig:badcenter} 
could appear.

Here we are using straight-line edge crossing reduction algorithms
for our confluent layered drawings with curve edges. 
Readers may suspect the equality of the crossing number in 
the straight-line drawing for the new three-layer graph $G'$
and the crossing number of our curve edge drawings. 
We will confirm this equality after we describe the
generation of curves in the next section.

\subsection{Curves}

After the positions of tree centers 
(and the positions of nodes if not given) are computed, 
we are now ready to place the confluent tracks for the edges.

We use \bez curves to draw the curve edges in confluent drawings.
Given a set of control points $P_1,P_2,\ldots,P_n$, the corresponding
\bez curve is given by 

\begin{equation}
C(u)=\sum_{k=0}^nP_k\,B_{k,n}(u)\qquad 0\le u \le 1\thinspace,\label{eqn:bcur}
\end{equation}

where $B_{k,n}(u)$ is a Bernstein polynomial
\begin{equation}
B_{k,n}(u)=\frac{n!}{k!\;(n-k)!}\;u^k(1-u)^{n-k}\thinspace.\label{eqn:bern}
\end{equation}

\bez curves have some nice properties that 
are suitable for our confluent tracks.
The first property is that a \bez curve always passes its
first and last control point.  
The second is that a \bez curve always stays within the convex hull
formed by its control points.
In addition,  the tangents of a \bez curve at the endpoints are
$P_1-P_0$ and $P_n-P_{n-1}$. 
Thus it is easy to connect two \bez curves while still
maintaining the first order continuity: just let $P_n=P'_0$ 
and let the control points
$P_{n-1}$,$P_n=P'_0$,and~$P'_1$ co-linear.

\begin{figure}[htbp]
  \psfrag{P0}{$P_0$} \psfrag{P1}{$P_1$} \psfrag{P2}{$P_2$} 
  \psfrag{P3=P0prime}{$P_3=P'_0$} \psfrag{P1p=P1pp}{$P'_1=P''_1$}
  \psfrag{P2p}{$P'_2$} \psfrag{P3p}{$P'_3$} 
  \psfrag{P2pp}{$P''_2$} \psfrag{P3pp}{$P''_3$} 
\vspace*{-7pt}
  \centering
  \includegraphics[height=2in]{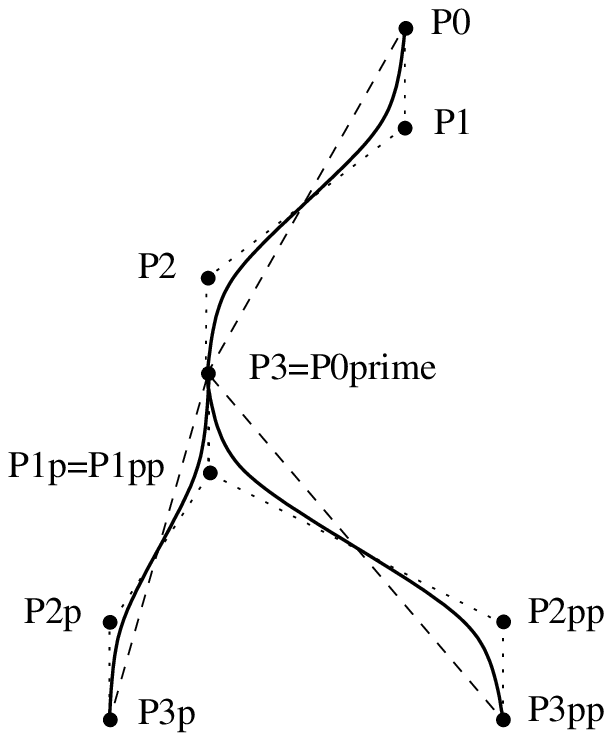}
\vspace*{-7pt}
  \caption{\bez curves}
  \label{fig:bcurves}
\end{figure}

The confluent track between each node and the tree center 
is drawn as a \bez curve.
In our program we use cubic \bez curves 
($n=4$ in Equation~\ref{eqn:bern}).  Each such a
curve has four control points, chosen as shown in
\figname\ref{fig:bcurves}. 

More formally, assume we are given the following input for a 
biclique $B_i$:
\journal{
\begin{itemize}
\setlength{\itemsep}{0pt}
\setlength{\parsep}{0pt}
\item $y_u$, the $y$-coordinate of the upper level.
\item $y_l$, the $y$-coordinate of the lower level.
\item $y_c$, the $y$-coordinate of the tree center level 
  (or just $\frac{1}{2}(y_u+y_l)$).
\item $x_i$, the $x$-coordinate of the tree center for $B_i$.
\item $x_{ij}$'s, the $x$-coordinates of nodes in biclique $B_i$.
\end{itemize}
}
\conferenceonly{$y_u$,$y_l$, and $y_c$ are the $y$-coordinates of 
the upper, lower, and tree center levels, respectively.
$x_i$ is the $x$-coordinate of the tree center for $B_i$. 
$x_{ij}$'s are the $x$-coordinates of nodes in biclique $B_i$.
}
Let $\Delta y$ be a distance parameter 
that controls the shape of the curve edges.
When node $j$ is in the upper level, 
the four control points are
$P_0=(x_{ij}, y_u)$, $P_1=(x_{ij}, y_u+\Delta y)$, 
$P_2=(x_i, y_c-\Delta y)$, and~$P_3=(x_i, y_c)$. 
When node $j$ is in the lower level, 
the four control points are
$P_0=(x_i, y_c)$, $P_1=(x_i, y_c+\Delta y)$, 
$P_2=(x_{ij}, y_l-\Delta y)$, and~$P_3=(x_{ij}, y_l)$. 

From Equation~\ref{eqn:bcur}, 
it is not hard to verify that 
in a confluent layered drawing, two
\bez curves cross each other 
if and only if the corresponding
straight-line edges (dashed lines in \figname\ref{fig:bcurves})
of the bicliques
cross each other, given that the control points are chosen 
as above.
This should clear the doubt that appears at the end of 
Section~\ref{sec:redu-xings}.

\section{Multi-depth Confluent Layered Drawings}
\label{sec:multi}

So far we have introduced the method of confluent layered drawings:
replacing subsets of edges in a two-layer graph by 
tree-like structures.
This method can be extended to obtain drawings 
that display richer information.  
The extended drawings are called 
\textit{multi-depth confluent layered drawings\/}.

The idea is as follows: 
after the biclique cover for a two-layer graph 
$G=(U,L,E)$ is computed, 
the tree center nodes are viewed as a middle layer $M$,
and a new three layer graph 
$G'=(U,M,L,E')$
is constructed as in 
Section~\ref{sec:redu-xings}. 
The same biclique cover algorithm is then applied to
$G'$ twice, 
once for the subgraph induced by $U\cup M$;
once for the subgraph induced by $M\cup L$.
By applying this approach recursively, 
we get biclique covers at different depth.
In the final drawing, only biclique covers 
at the largest depth are replaced by sets of tree-like structures.
The final drawing 
is a multi-depth confluent layered drawing.  
The drawings discussed before this section
are all \textit{depth-one (confluent layered) drawings\/}.

In a depth-one drawing, we compute a biclique cover 
and lay out the biclique cover.
In general, for a depth-$i$ drawing,
we need to compute $2^i-1$ biclique covers and 
$2^{i-1}$ biclique covers are laid out.

\begin{figure}[htbp]
  \centering
  \includegraphics[width=.42\textwidth]{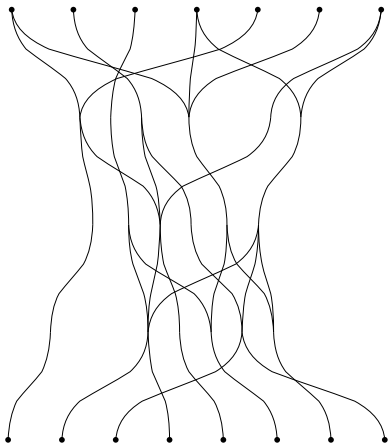}
  \caption{Depth-two confluent drawing (on same input as the drawing
    of \figname\ref{fig:first}).}
  \label{fig:depth-two}
\end{figure}

An example drawing of depth-two is shown in \figname\ref{fig:depth-two}.

Because the control points for the \bez curves 
are chosen in a way such that
the tangents at the endpoints of the \bez curves are all vertical,
it is guaranteed that
all segments of a path are connected seamlessly and smoothly
in multi-depth drawings.
Readers probably have already noticed some wavy edges
in the drawing of \figname\ref{fig:depth-two}.
It is because a single edge
biclique ($K_{1,1}$) is also drawn as two \bez curves.  
We offer an option in our program to do a simple treatment 
for these single edge bicliques: draw them as a single \bez
curves instead of two.  
But after this special treatment is applied, 
the crossing property is not preserved any more.  
That means 
two curve segments could have crossing(s), 
even though 
their corresponding edges in~$G'$
don't cross each other
in a straight-line drawing. 

Multi-depth drawings may further reduce the number of crossings.  
They also show a richer structure than the depth-one drawings, which
only display bicliques.  For example we can observe relationships
between bicliques in a depth-two confluent layered drawing.
However higher depth requires more 
computations of biclique covers, and generates more dummy centers.  The
former leads to the increasing of time and space complexity, while the
latter could result in a more complicate confluent drawing.  We feel
that drawings with depth higher than two are not very practically
useful. 

\section{Real-world Examples}
\label{sec:examples}

\begin{figure}[htb]
\vspace*{-7pt}
  \centering
  $${\includegraphics[width=\textwidth]{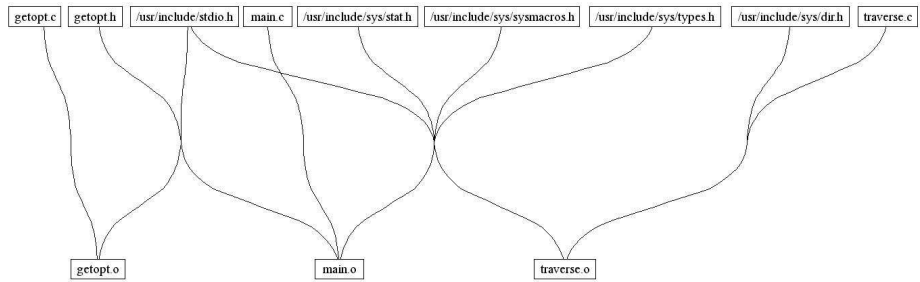}}\atop
    {\hbox{(a) ``Derives'' relation for the Shar program}}$$

  $${\includegraphics[width=0.75\textwidth]{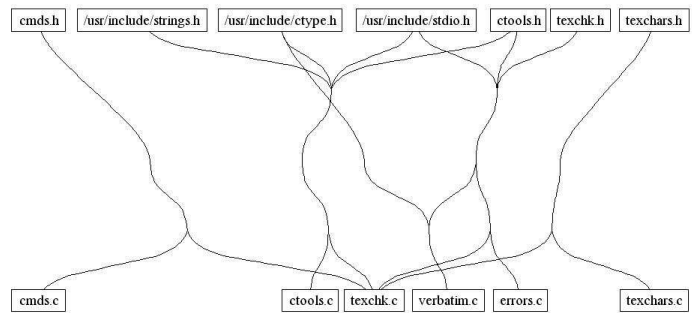}}\atop
    {\hbox{(b) ``Includes'' relation for the Texchk program}}$$
\vspace*{-7pt}
  \caption{Confluent drawing for examples of
  Newbery~\cite{n-ecmcd-89}.}
  \label{fig:newbery1}
\end{figure}

We list two example drawings of real-world graphs in 
\figname\ref{fig:newbery1}.  
We implemented the algorithm of computing biclique cover
using the RLF heuristic.  
For the center placement we implemented the barycenter method.
We assume that besides the two-layer graph,
the input also includes the positions of (fixed) nodes
in upper and lower levels 
(possibly output by other algorithms that take 
labels and other information into account.)
\conferenceonly{
The result drawing is written into a file of 
DOT format~\cite{kn-dgd-95}. 
The \texttt{neato} program in the 
Graphviz package~\cite{graphviz2001} 
is then used to generate the graphic file in
a desired format.
}
\journal{
Our program computes the biclique covers, 
the positions of tree centers, and the curve edges.
The result is then written into a file of 
DOT format~(\cite{kn-dgd-95}).
We choose DOT format because 
the \texttt{neato} program in
the Graphviz package~\cite{graphviz2001}
has a no-op option, which makes \texttt{neato} to respect
the node positions and the edge positions specified in the input. 
In other words, our program outputs the layout of the drawing 
into a DOT file, and \texttt{neato} only serves as a powerful
converter 
that transforms the layout into various graphic formats
such as PostScript, SVG, FIG, JPEG, etc.
}
\figname\ref{fig:newbery1}~(a) 
is a depth-one drawing.  
\figname\ref{fig:newbery1}~(b) is 
a depth-two drawing with 
the special smoothing treatment applied.

\section{Conclusions and Acknowledgments}
\label{sec:concl}

In this paper we introduce a new method 
-- confluent layered drawing,
for visualizing layered
graphs.  It combines the layered drawing 
technique with the relaxed confluent drawing approach.   
\journal{
We describe a 
reduction from biclique cover problem 
to a special kind of edge coloring
problem and use this reduction in our confluent layered drawing
algorithm.   We show how to lay out the bicliques. 
The idea can be extended to obtain multi-depth drawings.
Examples of real-world graphs are also given.  
}
There are still interesting open problems, e.g., 
how to test 
whether a layered graph has a crossing-free confluent layered drawing?
How to minimize the crossing of the drawing among all possible 
biclique covers?  It is also useful to investigate 
better ways for  
visualizing confluent tracks.

We would like to thank anonymous referees for their helpful comments.


\end{document}